\def   \ni {\noindent}

\def   \ssk {\vskip  5truept}

\def   \bsk {\vskip 15truept}

\def   \newline {\hfil\break}

\def\ref{\par\noindent\hangindent 20pt}
\def\mincir{\raise -2.truept\hbox{\rlap{\hbox{$\sim$}}\raise5.truept
\hbox{$<$}\ }}
\def\magcir{\raise -2.truept\hbox{\rlap{\hbox{$\sim$}}\raise5.truept
\hbox{$>$}\ }}

\documentstyle[epsfig]{article}
\begin{document}

\hsize 5truein
\vsize 8truein
\font\abstract=cmr8
\font\keywords=cmr8
\font\caption=cmr8
\font\references=cmr8
\font\text=cmr10
\font\affiliation=cmssi10
\font\author=cmss10
\font\mc=cmss8
\font\title=cmssbx10 scaled\magstep2
\font\alcit=cmti7 scaled\magstephalf
\font\alcin=cmr6 
\font\ita=cmti8
\font\mma=cmr8
\def\ref{\par\noindent\hangindent 15pt}
\null
%\vskip 3.0truecm
%\baselineskip = 12pt

% beginning of font "title"

\title{\ni 
EXTREME SYNCHROTRON BLAZARS IN OUTBURST: THE POTENTIALITY OF INTEGRAL 
IMAGING AND SPECTROSCOPY}

% beginning of font "author and affiliation"
\bsk \bsk
\author{\ni E.~Pian$^{1}$, G.~Malaguti$^{1}$, L.~Maraschi$^{2}$, 
G.~Ghisellini$^{3}$, G.~G.~C.~Palumbo$^{4,1}$}

\bsk
\affiliation{$^{1}$ ITESRE-CNR, Via Gobetti 101, 40129 Bologna, Italy}

\affiliation{$^{2}$ Osservatorio Astronomico di Brera, V. Brera 28, 20121 Milano,
Italy}

\affiliation{$^{3}$ Osservatorio Astronomico di Brera, V. Bianchi 46, 22055 Merate
(Lecco), Italy}

\affiliation{$^{4}$ Dipartimento di Astronomia di Bologna, V. Zamboni 33,
40126 Bologna, Italy}
                                                
\bsk
\baselineskip = 12pt

% beginning of font "abstract and keywords"
\abstract{ABSTRACT \ni

Only $\sim$10 blazars have been detected in soft gamma-rays by OSSE and the PDS,
often in flaring state. We investigate the impact of the imaging and spectral
capabilities of the IBIS instrument on board INTEGRAL on the study of blazars. The
objectives are: 1) to remove ambiguities in the identification of some blazars as
soft gamma-ray emitters;  2)  to determine the spectral shape of ``extreme
synchrotron" blazars, namely sources with high $F_X/F_{radio}$ ratios, for which
the synchrotron component peaks at far UV-soft X-ray frequencies. We expect that
in these sources the synchrotron peak shifts to higher energies during outbursts,
making them strong soft gamma-ray emitters, and therefore possible targets for
INTEGRAL wide band X- and gamma-ray spectroscopy.  }

\bsk
\baselineskip = 12pt

\bsk
\baselineskip = 12pt

% beginning of font "text"

\text{\ni 1. THE SOFT GAMMA-RAY EMISSION OF BLAZARS
\ssk
\ni     

%--------------------------  figure 1
%this section shows how to insert a figure in the text
\begin{figure}
%\vspace{-0.3cm}
\centerline{\psfig{file=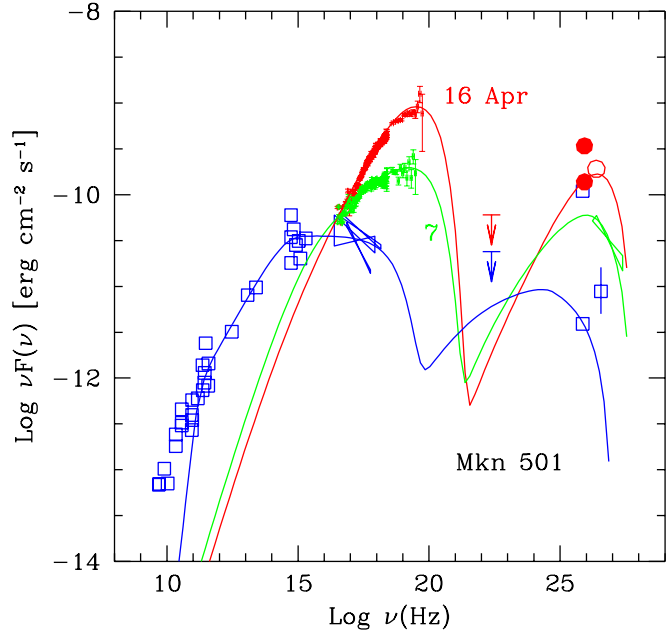, width=10cm}}
%\vspace{-0.2cm}
\caption{FIGURE 1. 
Spectral energy distribution of Mkn~501.  BeppoSAX data from 7 and 16 April
1997 are indicated as labeled.  Nearly simultaneous Whipple TeV data are
indicated as filled circles, while the open circle (13 April 1997) and the TeV
spectral fit (15-20 March 1997)  along with its 1-$\sigma$ confidence range are
from the HEGRA experiment.  Non-simultaneous measurements collected from the
literature are shown as open squares (see references in [6]). The
solid lines indicate fits with a one-zone, homogeneous synchrotron self-Compton
model. 
For all models the size of the emitting region is $R = 5\times 10^{15}$
cm, the beaming factor is $\delta = 15$ and the magnetic field is $B\sim 0.8$
Gauss. For the ``quiescent state'', the intrinsic luminosity (corrected for
beaming)  is $L^\prime = 4.6\times 10^{40}$ erg s$^{-1}$, and electrons are
continuously injected with a power-law distribution ($\propto \gamma^{-2}$)
between $\gamma_{min}=3\times 10^3$ and $\gamma_{max}=6\times 10^5$.  For the
fit to the 7 April spectrum, $L^\prime = 1.8\times 10^{41}$ erg s$^{-1}$ and
the injected electron distribution ($\propto \gamma^{-1.5}$)  extends from
$\gamma_{min} = 1\times 10^4$ to $\gamma_{max}=3\times 10^6$.  For the fit to
the 16 April spectrum, $L^\prime = 5.5\times 10^{41}$ erg s$^{-1}$ and the
injected electron distribution ($\propto \gamma^{-1}$) extends from
$\gamma_{min}=4\times 10^5$ to $\gamma_{max}=3\times 10^6$.  For the 7 and 16
April models, the seed photons for the Compton scattering are the sum of those
produced by the assumed electron distribution plus those corresponding to the
quiescent spectrum. 
%\vspace{-0.5cm}
}
\end{figure}

%---------------------------------

Blazars are radio-loud active galactic nuclei with highly luminous and highly
variable multiwavelength continua.  Their spectral energy distributions 
from the radio to the gamma-ray band can be described in a general way as composed
of two broad peaks in the power per decade, corresponding to the synchrotron and
inverse Compton emission, respectively (see e.g., [1]).  Blazars of
different luminosities appear to lie along a spectral sequence [2] whereby both peaks
fall at higher frequencies for lower luminosity objects. 
Thus highly luminous objects are ``red", meaning that the first peak falls at
frequencies smaller than the optical range, while low luminosity objects are
``blue", having peak frequency beyond the UV range.  For red blazars the second
spectral component peaks within or below the EGRET (0.1-10 GeV)  range, while for
blue blazars it peaks beyond the EGRET range.
The OSSE and PDS instruments on board CGRO and BeppoSAX, respectively, have
detected $\sim$10 blazars in the soft gamma-rays ($\sim$15--200 keV).  Among
these, most have rather flat spectra ($\alpha < 1$, with $F_\nu
\propto \nu^{-\alpha}$), indicating the inverse
Compton origin of radiation in this band [3-5].

For Mkn 501, during the X-ray outburst observed in April 1997 by BeppoSAX and RXTE
[6,7], both OSSE and the PDS measured a nearly
horizontal slope of the energy spectrum ($\alpha \simeq 1$), identified with the
peak of the synchrotron emission (Fig.~1).  This means that in this blue blazar,
the maximum of the synchrotron component had shifted during the outburst by nearly
three decades in energy, reaching the soft gamma-ray frequencies, and its high
energy tail might have extended to $\sim$1 MeV.  This suggests that for some
blazars in outburst, the soft gamma-ray radiation might be produced via the
synchrotron mechanism, instead of inverse Compton, thus implying the presence of
extremely energetic electrons.  A contribution from the inverse Compton component
might still be present toward higher energies ($\magcir 1$ MeV). 

\bsk
\ni 2. PLANNED IBIS OBSERVATIONS AND EXPECTED RESULTS
\ssk
\ni 

These sources, which have been also named ``Extreme Synchrotron Blazars" (ESB,
[8]), will have the characteristics of blue blazars in quiescent state.  Their
synchrotron peak is usually in the extreme UV/soft X-ray band, but it could shift
to higher energies during outbursts, namely when fresh, higher energy electrons
are injected, allowing their detection in hard X-rays and soft gamma-rays. The
ratio of their X-ray to radio output is high, if compared with that of red
blazars.  We have explored the possibility of doing spectroscopy of ESB in the
soft gamma-ray range with the IBIS instrument to be operating on board INTEGRAL.

%\bsk
%\ni
%3. EXPECTED RESULTS
%\ssk
%\ni

Fig.~2 reports the PDS spectra of three blazars compared with the IBIS sensitivity
curves at 3$\sigma$. Detection of the spectrum of an ESB of brightness and
spectrum comparable with those of Mkn 501, and assuming some downward curvature of
$\Delta \alpha \sim 1$ beyond 200 keV, will be possible up to $\sim$500 keV with a
significance $\ge 3 \sigma$.  We expect to observe with IBIS one or two ESBs in
outburst, as part of a Target-of-Opportunity program triggered by optical
monitoring. 
Determining with IBIS the soft gamma-ray spectrum of an ESB in the range 15-1000
keV will allow us to explore the highest energy end of the synchrotron spectrum,
and,
due to the wide energy range, to possibly disentangle the synchrotron and inverse
Compton components, and thus determine with high precision the maximum electron
energies reached during these powerful outbursts.  This will ultimately improve
our understanding of the acceleration mechanisms which produce these events.

%--------------------------  figure 2
%this section shows how to insert a figure in the text
\begin{figure}
\vspace{-0.6cm}
\centerline{\psfig{file=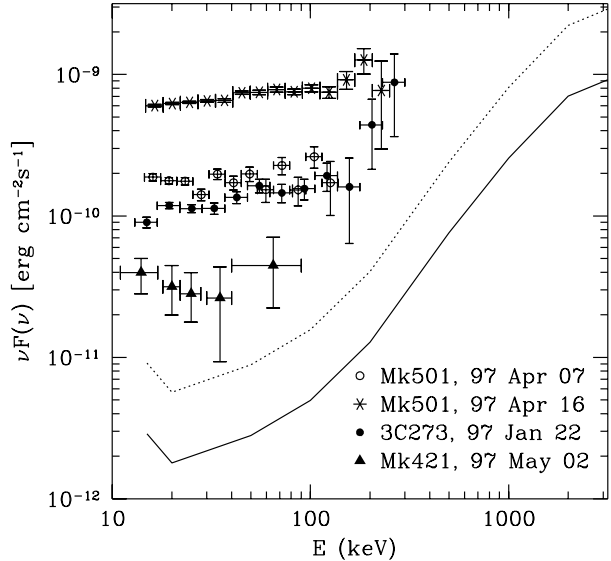, width=9cm}}
\vspace{-0.3cm}
\caption{FIGURE 2. 
PDS spectra of blazars observed with BeppoSAX (from [4-6].  
The sensitivity curves of IBIS
at 3$\sigma$, corresponding to an observing time of $10^5$ seconds (dotted) and of $10^6$
seconds (solid) are shown.
\vspace{-0.2cm}

}
\end{figure}

%---------------------------------

\bsk
\ni
3. THE PROBLEM OF THE CONTAMINATION
\ssk
\ni

The fine angular resolution of IBIS ($12^{\prime}$)  will allow us to
disentangle the soft gamma-ray flux of some sources which might be contaminated
by nearby objects in OSSE detections, and possibly also in the PDS observations. 
For instance, the BL Lac object PKS~2155--304, one of the best studied blazars
at all frequencies, lies at $\sim 1.8^{\circ}$ from the Seyfert 2 galaxy
NGC~7172, of comparable average brightness. OSSE (field of view $4^{\circ}
\times 11^{\circ}$) observations could not distinguish the flux from either
source [3], while the PDS has a sufficiently small field of view ($1.3^{\circ}$
at FWHM) to exclude NGC~7172 when pointing at PKS~2155--304. Another source, the
quasar PKS~2149--306, which has a hard X-ray spectrum [9] and an extrapolated
soft gamma-ray brightness similar to, or larger than that of PKS~2155--304, lies
at$\sim 1.5^{\circ}$, and might therefore marginally contaminate the PDS
response.  Although in the case of  PKS 2155--304 contamination of both flux and
background PDS measurements is not severe (not exceeding $\sim$5\%),
it might be a concern for other sources. The imaging capabilities of IBIS 
will completely avoid this problem.
}

%\bsk
%\baselineskip = 12pt
%{\abstract \ni ACKNOWLEDGMENTS
%
%}

\bsk
\baselineskip = 12pt

% beginning of font "references"

{\references \ni REFERENCES
\ssk

[1] Ulrich, M.-H., Maraschi, L., \& Urry, C. M. 1997, ARA\&A, 35, 445;
[2] Fossati, G., et al. 1998, submitted to MNRAS (astro-ph/9804103);
[3] McNaron-Brown, K., et al. 1995, ApJ, 451, 575;
[4] Haardt, F., et al. 1998, A\&A, in press (astro-ph/9806229);
[5] Fossati, G., et al. 1998, in ``The Active X-ray Sky: Results from BeppoSAX 
    and Rossi-XTE", Nuclear Physics B (Proc. Suppl.) 69/1-3, eds. L. Scarsi, H.
    Bradt, P. Giommi and F. Fiore, Elsevier, p. 423;
[6] Pian, E., et al. 1998, ApJ, 492, L17;
[7] Catanese, M., et al. 1997, ApJ, 487, L143;
[8] Ghisellini, G. 1998, in ``The Active X-ray Sky: Results from BeppoSAX 
    and Rossi-XTE", Nuclear Physics B (Proc. Suppl.) 69/1-3, eds. L. Scarsi, H.
    Bradt, P. Giommi and F. Fiore, Elsevier, p. 397;
[9] Cappi, M., et al. 1997, ApJ, 478, 492.
}

\end{document}